\documentclass[twocolumn,superscriptaddress,aps,preprintnumbers,amsmath,amssymb,sort&compress,nofootinbib
,prl
]{revtex4-2}
  
\usepackage{amsfonts,amsmath,amssymb,ascmac,bm,tensor}
\usepackage{fnpct} 
\usepackage{comment}
\usepackage{ifpdf}
\usepackage{slashed}
\usepackage{color}
\usepackage[mathscr]{eucal}
\usepackage{fourier}
\usepackage[utf8]{inputenc}
\usepackage{physics}
\usepackage{cancel}

\ifpdf
  \usepackage{graphicx}     
  \usepackage[bookmarksopen,colorlinks=true,linkcolor=bblue,citecolor=bblue,urlcolor=ppink]{hyperref}
\else     
\fi

\definecolor{red}{rgb}{1,0,0}
\definecolor{darkred}{rgb}{0.6,0,0}
\definecolor{darkgreen}{rgb}{0.992447,0.623778,0.034597}
\definecolor{ppink}{rgb}{1,0.4,0.4}
\definecolor{bblue}{rgb}{0.284602,0.317763,0.963947}
\definecolor{purple}{rgb}{0.5 ,0, 0.7}

\newcommand{\DD}{\text{D}}
\newcommand{\ee}{\mathrm{e}}

\makeatletter
\newcommand\footnoteref[1]{\protected@xdef\@thefnmark{\ref{#1}}\@footnotemark}
\makeatother
 
\allowdisplaybreaks[1]

\begin{document}


\title{
NANOGrav results and LIGO-Virgo primordial black holes in axion-like curvaton model
}

\author{Keisuke Inomata}
\affiliation{Kavli Institute for Cosmological Physics, The University of Chicago, Chicago, IL 60637, USA}
\author{Masahiro Kawasaki}
\affiliation{ICRR, University of Tokyo, Kashiwa, 277-8582, Japan}
\affiliation{Kavli IPMU (WPI), UTIAS, University of Tokyo, Kashiwa, 277-8583, Japan}
\author{Kyohei Mukaida}
\affiliation{CERN, Theoretical Physics Department, CH-1211 Geneva 23, Switzerland}
\affiliation{DESY, Notkestra{\ss}e 85, D-22607 Hamburg, Germany}
\author{Tsutomu T.~Yanagida}
\affiliation{T. D. Lee Institute and School of Physics and Astronomy, Shanghai Jiao Tong University, 800 Dongchuan Rd, Shanghai 200240, China}
\affiliation{Kavli IPMU (WPI), UTIAS, University of Tokyo, Kashiwa, 277-8583, Japan}

\begin{abstract}
\noindent
We discuss a possible connection between the recent NANOGrav results and the primordial black holes (PBHs) for the LIGO-Virgo events.
In particular, we focus on the axion-like curvaton model, which provides a sizable amount of PBHs and GWs induced by scalar perturbations around the NANOGrav frequency range.
The inevitable non-Gaussianity of this model suppresses the induced GWs associated with PBHs for the LIGO-Virgo events to be compatible with the NANOGrav results.
We show that the axion-like curvaton model can account for PBHs for the LIGO-Virgo events and the NANOGrav results simultaneously.
\end{abstract}

\date{\today}
\maketitle
\preprint{CERN-TH-2020-182}
\preprint{DESY 20-188}

\emph{Introduction.}---
The NANOGrav collaboration has recently reported the results of the 12.5-year pulsar timing observation, which shows strong evidence of a stochastic process around $\mathcal O(\text{nHz})$ with a common amplitude and a common spectral index across all pulsars~\cite{Arzoumanian:2020vkk}. 
Although quadrupolar spatial correlations of the stochastic process, which should exist for gravitational wave (GW) signals, have not been found yet, it is worth investigating potential implications of the process in terms of stochastic GWs.

In this letter, we discuss a potential connection between the NANOGrav results and the primordial black holes (PBHs)~\cite{Hawking:1971ei,Carr:1974nx,Carr:1975qj}.
PBHs have recently attracted a lot of attention as candidates of dark matter (DM)~\cite{Carr:2016drx,Inomata:2017okj,Inomata:2017vxo} and the black holes (BHs) detected by the LIGO-Virgo collaboration~\cite{Bird:2016dcv,Clesse:2016vqa,Sasaki:2016jop} (see also Refs.~\cite{Sasaki:2018dmp,Carr:2020gox,Green:2020jor} for recent reviews).
As a probe of PBHs, GWs induced by scalar perturbations are often discussed because a sizable amount of PBHs require large-amplitude scalar perturbations on small scales, which makes the GWs large enough to be investigated by the current and future GW experiments~\cite{Ananda:2006af,Baumann:2007zm,Saito:2008jc,Saito:2009jt,Inomata:2016rbd,Ando:2017veq,Espinosa:2018eve,Kohri:2018awv,Cai:2018dig,Bartolo:2018evs,Bartolo:2018rku,Unal:2018yaa,Byrnes:2018txb,Inomata:2018epa,Clesse:2018ogk,Cai:2019amo,Cai:2019jah,Wang:2019kaf,Cai:2019elf,Zhou:2020kkf}. 
In particular, the LIGO-Virgo PBHs with $\sim \mathcal O(10) M_\odot$ predict the large GWs around $\mathcal O(\text{nHz})$, which is close to the NANOGrav frequency.

There are already some works discussing the NANOGrav results in the context of the induced GWs associated with PBHs.
Ref.~\cite{Vaskonen:2020lbd} showed that the NANOGrav signal is too small to be compatible with the induced GWs associated with PBHs for the LIGO-Virgo events in their setups, but on the other hand it can be consistent with the PBHs as the primordial seeds of supermassive BHs.
Also, possible connections of the NANOGrav results with the PBHs for DM~\cite{DeLuca:2020agl}, $\mathcal O(1)M_\odot$ BHs~\cite{Kohri:2020qqd}, and planet-mass BHs~\cite{Domenech:2020ers} were discussed (see also Refs.~\cite{Sugiyama:2020roc,Bhattacharya:2020lhc}).

The main aim of this letter is to show that the axion-like curvaton model~\cite{Kawasaki:2012wr,Ando:2017veq,Ando:2018nge} can generate PBHs for the LIGO-Virgo events that are consistent with the NANOGrav results.
The key difference from the setups in Ref.~\cite{Vaskonen:2020lbd} is that we take into account the primordial non-Gaussianity of curvature perturbations, which is inevitably produced in the curvaton model.
The primordial non-Gaussianity suppresses the induced GWs with the PBH abundance fixed, which enables our model to explain PBHs for the LIGO-Virgo events and the NANOGrav results simultaneously.

\emph{Axion-like curvaton model.}---
First, we summarize the basic properties of the axion-like curvaton model (see Refs.~\cite{Kawasaki:2013xsa,Ando:2017veq,Ando:2018nge} for detail).
This model was originally introduced in the framework of supersymmetry~\cite{Kasuya:1996ns,Kasuya:2009up}.
The relevant field corresponds to the flat direction, denoted as $\Psi = \varphi \ee^{i \theta}/\sqrt{2}$.
After taking into account the supergravity effect, we can approximate the effective potential for $\varphi$ as
\begin{align}
	V(\varphi) = \frac{1}{2}c H^2(\varphi - f)^2,
	\label{eq:v_varphi_pot}
\end{align}
where $H$ is the Hubble parameter during the inflation, $c$ is the coefficient coming from the supergravity effect, and $f$ is a symmetry breaking scale.
We assume that $c$ is of $\mathcal O(1)$ and the initial value of $\varphi$ is much larger than $f$, which makes $\varphi$ roll down the potential to the minimum $f$ during inflation.

Once $\varphi$ reaches the minimum $f$, the curvaton is given by $\sigma \equiv \theta f$.
Similarly to the QCD axion, we assume that a certain non-perturbative dynamics generates a mass to the curvaton at some point after reheating.
Its potential is given by
\begin{align}
	V(\sigma) = \Lambda^4 \left[ 1 - \cos\left( \frac{\sigma}{f} \right) \right] \simeq \frac{1}{2} m_\sigma^2 \sigma^2,
\end{align}
where $\Lambda$ is the scale of the non-perturbative dynamics and the curvaton mass is obtained from $m_\sigma \equiv \Lambda^2/f$. 
The second equality is valid when the curvaton is near its minimum.
Once the curvaton acquires its mass, it starts to oscillate around the minimum and behaves as the non-relativistic matter component.
During a radiation-dominated (RD) era, its energy fraction in the total energy continues to grow.
In this work, we consider the case where the curvaton decays to radiation before it dominates the Universe.

In the axion-like curvaton model, the power spectrum of curvature perturbations on small scales, relevant to the PBH production, can be approximated by 
\begin{align}
	\mathcal P_\zeta(\eta,k) = \begin{cases}
	 A_\zeta(\eta) \left( \frac{k}{k_*} \right)^{n_1 -1} & \text{for } k < k_*\\
	 A_\zeta(\eta) \left( \frac{k}{k_*} \right)^{n_2 -1} & \text{for } k \geq k_*\\	 
	\end{cases},
	\label{eq:pzeta_curvaton}
\end{align}
where $\eta$ is the conformal time and $k_*$ is the inverse of the horizon scale at the arrival of $\varphi$ to the minimum $f$.
Note that, on large scales observed by CMB, the power spectrum is dominated by the inflaton fluctuations and hence different from Eq.~\eqref{eq:pzeta_curvaton}.
$A_\zeta$ is given by 
\begin{align}
	A_\zeta(\eta) = \left( \frac{2r}{4 + 3 r} \right)^2  \left( \frac{H|_{\eta = 1/k_*}}{2 \pi f \theta} \right)^2,
\end{align}
where $H|_{\eta = 1/k_*}$ denotes its value at $\eta = 1/k_*$ and $\theta$ is the misalignment angle of the curvaton.
$r$ stands for the ratio between the energy densities of the curvaton and radiation until the curvaton decay, and it is frozen after its decay, \textit{i.e.},
\begin{align}
	r(\eta) = \begin{cases}
	r_\text{D} \frac{\eta}{\eta_\text{D}} \quad &\text{for}\  \eta < \eta_\DD \\
	r_\text{D} \quad & \text{for} \ \eta \geq \eta_\DD
	\end{cases},
\end{align}
where the subscript D indicates the value at the curvaton decay.
The tilt $n_{1}$ is related to the coefficient $c$ in Eq.~\eqref{eq:v_varphi_pot}: 
\begin{align}
	n_1 -1 = 3 -3\sqrt{1 - \frac{4}{9}c},
\end{align}
which originates from the evolution of the radial direction $\varphi$ during the inflation. 
Since a typical size of fluctuations of the phase-direction field at the horizon exit is $H/(2\pi)$ regardless of the value of $\varphi$ at that time, the fluctuation of the misalignment angle is given by $\delta \theta = H/(2\pi \varphi_\text{exit})$, with $\varphi_\text{exit}$ being the value at the horizon exit.
The curvaton perturbation, $\delta \sigma = f \delta \theta$, gets more suppressed for a larger value of $\varphi_\text{exit}$ corresponding to perturbations at a larger scale, \textit{i.e.}, a smaller $k$.
On the other hand, the other tilt $n_2$ stems from the time dependence of $H$, which was not taken into account in our previous work~\cite{Ando:2017veq}.
As we consider slow-roll inflation, $n_2 -1$ should be smaller than unity.
Note that $n_2$ is the tilt of the small-scale perturbations and therefore can be different from that on larger scales, determined by the Planck collaboration~\cite{Aghanim:2018eyx}.\footnote{
Using the slow-roll parameter $\epsilon = -\dot{H}/H^2$, $n_2 -1$ is written as $n_2-1 = -2\epsilon$.
If the spectral index is almost constant between $k_*$ and $k_D$,
the corresponding change of the inflaton field $\Delta I$ is given by $\Delta I \simeq \sqrt{1-n_2}M_\text{Pl}\,\log(k_D/k_*)$ ($M_\text{Pl}$: the Planck scale).
Thus, a significant tilt ($n_2 \lesssim 0.8$) implies large field inflation.
}

\emph{PBH abundance.}---
Next, we introduce equations connecting the power spectrum and the PBH abundance.
The scale of the perturbation for the PBH production is related to the PBH mass as~\cite{Inomata:2017vxo}, 
\begin{align}
	\label{eq:mpbh_k_rel}
	M_\text{PBH} \simeq& M_\odot \left( \frac{\gamma}{0.2}\right) \left( \frac{g}{10.75} \right)^{-1/6} \left( \frac{k}{1.9 \times 10^6\,\text{Mpc}^{-1}} \right)^{-2} \\
	\simeq &
	 M_\odot \left( \frac{\gamma}{0.2}\right) \left( \frac{g}{10.75} \right)^{-1/6} \left( \frac{f}{2.9\times10^{-9} \,\text{Hz}} \right)^{-2},
\end{align}
where $\gamma$ is the fraction of the PBH mass in the horizon mass at the production and $g$ is the effective degrees of freedom of radiation at that time.
As fiducial values, we take $\gamma = 0.2$~\cite{Carr:1975qj} and $g= 10.75$.
We do not include the effect of the critical collapse~\cite{Niemeyer:1997mt,Niemeyer:1999ak,Shibata:1999zs,Musco:2004ak} for simplicity because Ref.~\cite{Yokoyama:1998xd} implies that it does not modify the PBH abundance so much.

We adopt the Press-Schechter formalism throughout this work.\footnote{
	We leave the analysis based on the peaks theory for future work.
	See Refs.~\cite{Suyama:2019npc,Germani:2019zez,Tokeshi:2020tjq} for recent updates of the peaks theory for PBH mass spectra.
}
In this formalism, the production rate of PBHs at the time of formation, $\beta$, is given by
\begin{align}
	\beta(M) = \int_{\delta_\text{c}} \dd \delta \frac{1}{\sqrt{2 \pi} \sigma(M)} \exp\left( -\frac{\delta^2}{2  \sigma^2(M)} \right),
	\label{eq:beta_def}
\end{align}
where $\delta_\text{c}$ is the threshold of the PBH formation.
Here, we have assumed the perturbations follow the Gaussian statistics and we will explain how to take into account the non-Gaussianity later.
The variance of the smoothed density contrast is denoted by $\sigma^2$. 
(Hereafter, it does not indicate the curvaton field.)
The variance is obtained from
\begin{align}
	\sigma^2(M) = \int^\infty_0 \frac{\dd q}{q} \tilde W^2(q;R) D^2(qR) \frac{16}{81} (qR)^4 \mathcal P_\zeta(\eta = R,q),
\end{align}
where $M$ denotes the PBH mass, $R = 1/k(M)$ is the smoothing scale for the PBH production of mass $M$, $\tilde W$ is a window function in Fourier space, and $D(x)$ is the transfer function of the gravitational potential, $\Phi$, with the normalization of $\Phi = -\frac{3}{2} \zeta$ in the superhorizon limit.
Remember that the wavenumber is related to the PBH mass through Eq.~(\ref{eq:mpbh_k_rel}).
We take the following transfer function regardless of the scales:
\begin{align}
	D(x) = \frac{9}{x^2} \left[ \frac{\sin(x/\sqrt{3})}{x/\sqrt{3}} - \cos(x/\sqrt{3}) \right].
	\label{eq:transfer_rd}
\end{align}
Although the transfer function for $k \gtrsim k_\text{D} (\equiv 1/\eta_\text{D})$ could be modified~\cite{Ando:2017veq}, we expect such modification does not change our results so much because the main contribution comes from the scales of $k_* < k < k_\text{D}$.
In this letter, we take the real-space top-hat window function as a fiducial example and take $\delta_\text{c} = 0.51$ for a pure RD era~\cite{Young:2019osy} (see also Ref.~\cite{Ando:2018qdb} for the discussion on the window function dependence).
We also take into account a small modification of $\delta_\text{c}$ during the QCD phase transition based on the results in Ref.~\cite{Byrnes:2018clq}.
Note that, with the real-space top-hat window function, $\sigma^2(M) \simeq \mathcal P_\zeta(k)$ holds for a scale-invariant spectrum~\cite{Ando:2017veq,Ando:2018qdb}.
The current fraction of PBHs in DM over a logarithmic interval is 
\begin{align}
	f_\text{PBH}(M) \equiv \frac{1}{\Omega_\text{DM} } \frac{\dd \Omega_\text{PBH}(M)}{\dd \text{ln}\, M} \simeq \left( \frac{\beta(M)}{1.84 \times 10^{-8}} \right) \left( \frac{M}{M_\odot} \right)^{-1/2},
\end{align}
where $\Omega_\text{DM}$ is the current energy density parameter of DM and $\Omega_\text{PBH}(M)$ is the density parameter of PBHs whose masses are smaller than $M$.

\emph{Induced gravitational waves.}---
Here, we briefly review the fundamental equations of GWs induced by scalar perturbations.
We take the conformal Newtonian gauge
\begin{align}
  \dd s^2 = &a^2 \left[ -\left(1+2\Phi \right) \dd \eta^2 
  + \left( \left(1-2\Psi \right) \delta_{ij} + \frac{1}{2} h_{ij} \right) \dd x^i \dd x^j \right],
\end{align}
where $a$ is the scale factor, $\Phi$ and $\Psi$ are scalar perturbations, and $h_{ij}$ is the tensor perturbation corresponding to GWs.
Since we focus on the early Universe, we may assume the perfect fluid condition, $\Phi = \Psi$.
See Refs.~\cite{Hwang:2017oxa,Gong:2019mui,Tomikawa:2019tvi,DeLuca:2019ufz,Inomata:2019yww,Yuan:2019fwv,Giovannini:2020qta,Lu:2020diy,Chang:2020tji,Ali:2020sfw,Chang:2020iji,Chang:2020mky} for the recent discussion on the gauge (in)dependence of the induced tensor perturbations.

The energy density parameter of the induced GWs is
\begin{align}
\Omega_{\rm{GW}} (\eta, k) 
&=
\frac{1}{24} \left( \frac{k}{\mathcal H(\eta) } \right)^2 
\overline{\mathcal P_{h} (\eta,k)} ,
\label{eq:gw_formula}
\end{align}
where $\mathcal H$ is the conformal Hubble parameter.
$\overline{\mathcal P_{h}}$ is the time-averaged power spectrum of the induced GWs, given by~\cite{Ando:2017veq}
\begin{align}
\overline{\mathcal P_{h} (\eta,k)} = &\frac{1}{4} \int^\infty_0 \dd v \int^{1+v}_{|1-v|} \dd u \left[ \frac{4v^2 - (1+v^2 - u^2)^2}{4vu}  \right]^2 \nonumber \\
& \times \overline{I^2(u,v,k,\eta) } \left( \frac{4+3 r_\DD}{r_\DD} \right)^4 \mathcal P_\zeta(\eta_\DD,ku) \mathcal P_\zeta(\eta_\DD,kv).
\label{eq:p_h_formula}
\end{align}
The integrand $I(u,v,k,\eta)$ is obtained from
\begin{align}
I(u,v,k,\eta) =& \int^{x}_0 \dd \bar{x} \frac{a(\bar \eta)}{a(\eta)} k G_k(\eta, \bar \eta) F(ku,kv,\bar \eta),
\label{eq:i_formula_r}
\end{align}
where $x = k \eta$ and $\bar x = k\bar \eta$.
Since the curvaton energy is assumed to be subdominant in the Universe in our scenario, we may use the Green function in a RD era, namely
\begin{align}
k G_k(\eta, \bar \eta) = \sin[k(\eta - \bar \eta)] \Theta(\eta - \bar \eta),
\end{align}
where $\Theta(x)$ is the Heaviside step function.
The function $F$ in Eq.~\eqref{eq:i_formula_r} can be expressed by means of the transfer function of the gravitational potential, $T(\eta, k)$:
\begin{align}
\label{eq:f_def}
&F(ku,kv,\bar{\eta})= 4 \left[ 3 T(\bar \eta, k u)T(\bar \eta, kv) \right. \\  
&\;\;\; + \left. 2 \mathcal H^{-1}(\bar \eta) T'(\bar \eta, ku)T( \bar \eta, kv)  
+  \mathcal H^{-2}(\bar \eta) T'(\bar \eta, ku)T'( \bar \eta, kv) \right],  \nonumber
\end{align}
where the prime means a derivative with respect to $\bar \eta$.  
A concrete expression of $T(\eta, k)$ is given in Sec.IV of Ref.~\cite{Ando:2017veq}.
For scales that enter the horizon after the curvaton decay, $T(\eta,k)$ is almost the same as $D(k \eta)$, given in Eq.~\eqref{eq:transfer_rd}, except for the normalization.
On the other hand, for scales that enter the horizon before the curvaton decay, we need to pay attention to some issues overlooked in our previous work~\cite{Ando:2017veq}.
In the previous work, we implicitly assumed that the curvaton decay occurs instantaneously, and impose continuity on $T$ and $T'$ before and after the curvaton decay.
However, in a realistic situation, the curvaton decay occurs gradually on a time scale of its decay rate and such an approximation might not be valid, given the results in Ref.~\cite{Inomata:2019zqy}.
Although the situation of this work is not exactly the same as that of Ref.~\cite{Inomata:2019zqy}, some suppression of gravitational potential during the curvaton decay might occur similarly.
The study of this effect is beyond the scope of this letter and we use the expression of $T(\eta, k)$ in Ref.~\cite{Ando:2017veq} 
for simplicity.
For this reason, we should keep in mind that the GWs spectrum of this work could possibly be more suppressed on scales $k>k_\text{D}$.

During a RD era, the density parameter in Eq.~\eqref{eq:gw_formula} finally asymptotes to a constant value after the scalar perturbations on the peak scale ($k\gtrsim k_*$) enter the horizon because the induced GWs behave as radiation.
Taking into account the subsequent matter-dominated and dark-energy-dominated era, we obtain the following expression of the density parameter at present~\cite{Ando:2017veq}:
\begin{align}
	\Omega_\text{GW}(\eta_0, k)  = 0.83 \left( \frac{g_\text{const}}{10.75} \right)^{-1/3} \Omega_{\text{r},0} \Omega_\text{GW}(\eta_\text{const},k),
\end{align}
where $\eta_\text{const}$ is the conformal time when $\Omega_\text{GW}$ becomes constant at the RD era, $g_\text{const}$ is degrees of freedom at $\eta_\text{const}$, and $\Omega_{\text{r},0}$ is the current radiation density parameter.

\emph{Non-Gaussianities.}---
In the axion-like curvaton model, the curvature perturbations that produce PBHs inevitably have the primordial non-Gaussianity.
Besides, the non-linear relation between the density perturbations and the curvature perturbations induces the intrinsic non-Gaussianity for the density perturbations~\cite{Kawasaki:2019mbl,DeLuca:2019qsy,Young:2019yug}. 
The non-Gaussianity is characterized by 
\begin{align}
	\delta(\bm x) = \delta_\text{g}(\bm x) + \frac{\mu_3}{6 \sigma } (\delta^2_\text{g}(\bm x) - \sigma^2),
	\label{eq:delta_fnl}
\end{align}
where $\delta_\text{g}$ follows the Gaussian statistics and $\sigma^2$ is its variance.
Using the equations in Ref.~\cite{Kawasaki:2019mbl}, we numerically evaluate the skewness $\mu_3$ as 
\begin{align}
	\frac{\mu_3}{\sigma} \simeq p(n_\text{s}) f_\text{NL} - \frac{9}{4},
	\label{eq:skewness_exp}
\end{align}
where the first term is the contribution from the primordial non-Gaussianity and the second term is from the intrinsic non-linear relation.
The coefficient $p(n_\text{s})$ depends on the tilt of the power spectrum of curvature perturbation, e.g., $p(1) \simeq 3.8$ (scale invariant) and $p(0.8) \simeq 4.4$ (red-tilted, used later).
In the curvaton model, $f_\text{NL}$ is related to $r_\text{D}$ as 
\begin{align}
	f_\text{NL} = \frac{5}{12} \left( -3 + \frac{4}{r_\DD} + \frac{8}{4 + 3 r_\DD} \right).
\end{align}

The probability distribution function (PDF) of the Gaussian part is given by
\begin{align}
	P_\text{G} (\delta_\text{g}) = \frac{1}{\sqrt{2\pi} \sigma } \exp \left( - \frac{\delta_\text{g}^2}{2\sigma^2} \right).
\end{align}
Solving Eq.~(\ref{eq:delta_fnl}) with respect to $\delta_\text{g}$, we obtain
\begin{align}
	\delta_{\text{g} \pm}(\delta) = \frac{3 \sigma}{\mu_3} \left[ -1 \pm \sqrt{1+ \frac{2 \mu_3}{3} \left( \frac{ \mu_3 }{6} + \frac{\delta}{\sigma} \right)} \right].
\end{align}
The PDF of the non-Gaussian part is now written as
\begin{align}
	P_\text{NG}(\delta) = \sum_{i= \pm} \left| \frac{\dd \delta_{g,i}(\delta)}{\dd \delta} \right|  P_\text{G}(\delta_{\text{g},i}(\delta)).
\end{align}
Then, we can express $\beta$ as 
\begin{align}
	\beta =& \int_{\delta_\text{c}} \dd \delta \, P_\text{NG}(\delta) \nonumber \\
	=& \begin{cases}
	\int^{\infty}_{\delta_{\text{g} +} (\delta_\text{c})} P_\text{G}(\delta_\text{g}) \dd \delta_\text{g} 
	+ \int^{\delta_{\text{g} -}(\delta_\text{c})}_{-\infty} P_\text{G}(\delta_\text{g}) \dd \delta_\text{g} \ & \text{for} \ \ \mu_3 > 0 \\
	\int^{\delta_{\text{g} -} (\delta_\text{c})}_{\delta_{\text{g} +}(\delta_\text{c})} P_\text{G}(\delta_\text{g}) \dd \delta_\text{g}  \ & \text{for} \  \ \mu_3 < 0
	\end{cases}.
	\label{eq:beta_png}
\end{align}
Since $\Omega_\text{GW}$ is roughly proportional to $\mathcal P_\zeta^2$, we define the suppression factor of $\Omega_\text{GW}$ as 
\begin{align}
	Q \equiv \left( \frac{\mathcal P_\zeta(\eta_\DD,k_*)}{\mathcal P_\zeta^{\mu_3 = 0}(\eta_\DD,k_*)} \right)^2.
\end{align}
Here $\mathcal P_\zeta^{\mu_3 = 0}$ is the power spectrum with $\mu_3 = 0$ which gives the same $\beta$ as that for $\mathcal P_\zeta$ with non-zero $\mu_3$ in the numerator.
We take into account the non-Gaussianity effect on the induced GWs multiplying this suppression factor to the $\Omega_\text{GW}$ that is calculated with $\mathcal P_\zeta^{\mu_3 = 0}$.
Note that the non-Gaussianity also modifies Eq.~(\ref{eq:p_h_formula}) itself by $\mathcal O(\mathcal P_\zeta^4 f_\text{NL}^2)$, shown in Refs.~\cite{Cai:2018dig,Unal:2018yaa}.
Although this modification affects the shape of the GW spectrum in $f_\text{NL} \gg 1$, we consider the case of $f_\text{NL} \sim \mathcal O(1)$ in the following and therefore can safely neglect the modification.

\emph{Results.}---
As a fiducial example, we take the following parameter values:
\begin{align}
	&\mathcal P_\zeta^{\mu_3 = 0}(\eta_\DD,k_*) = 0.0067,\  k_*= 5\times 10^5\, \text{Mpc}^{-1},\,\nonumber \\
	& k_\text{D} = 1\times 10^8 \, \text{Mpc}^{-1}, \ n_1 = 2.5, \  n_2 = 0.8.
	\label{eq:parameter_set}
\end{align}

\begin{figure}[tbh] 
        \centering \includegraphics[width=1\columnwidth]{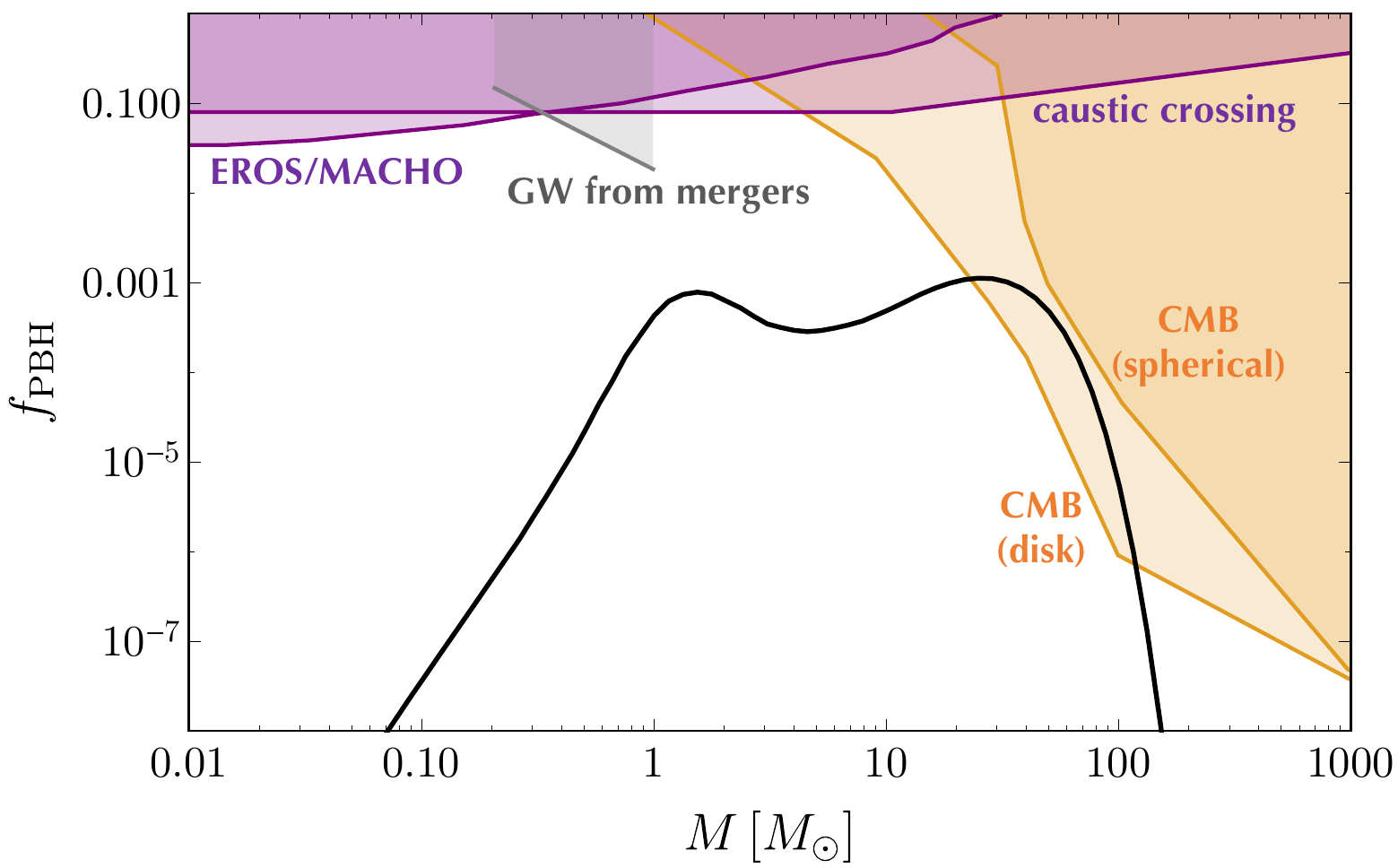}
        \caption{
        PBH mass spectrum (black solid) with the parameter set \eqref{eq:parameter_set}.
        The shaded regions are constrained by EROS/MACHO~\cite{Allsman:2000kg,Tisserand:2006zx}, caustic crossing events~\cite{Oguri:2017ock}, CMB anisotropy with the assumption of the spherical and the disk accretion onto PBHs~\cite{Serpico:2020ehh}, and GWs from mergers~\cite{Authors:2019qbw}.
        See also Ref.~\cite{Carr:2020gox} for other constraints.
        }
        \label{fig:pbh_mass_spectrum}
\end{figure}

\begin{figure}[tbh] 
        \centering \includegraphics[width=1\columnwidth]{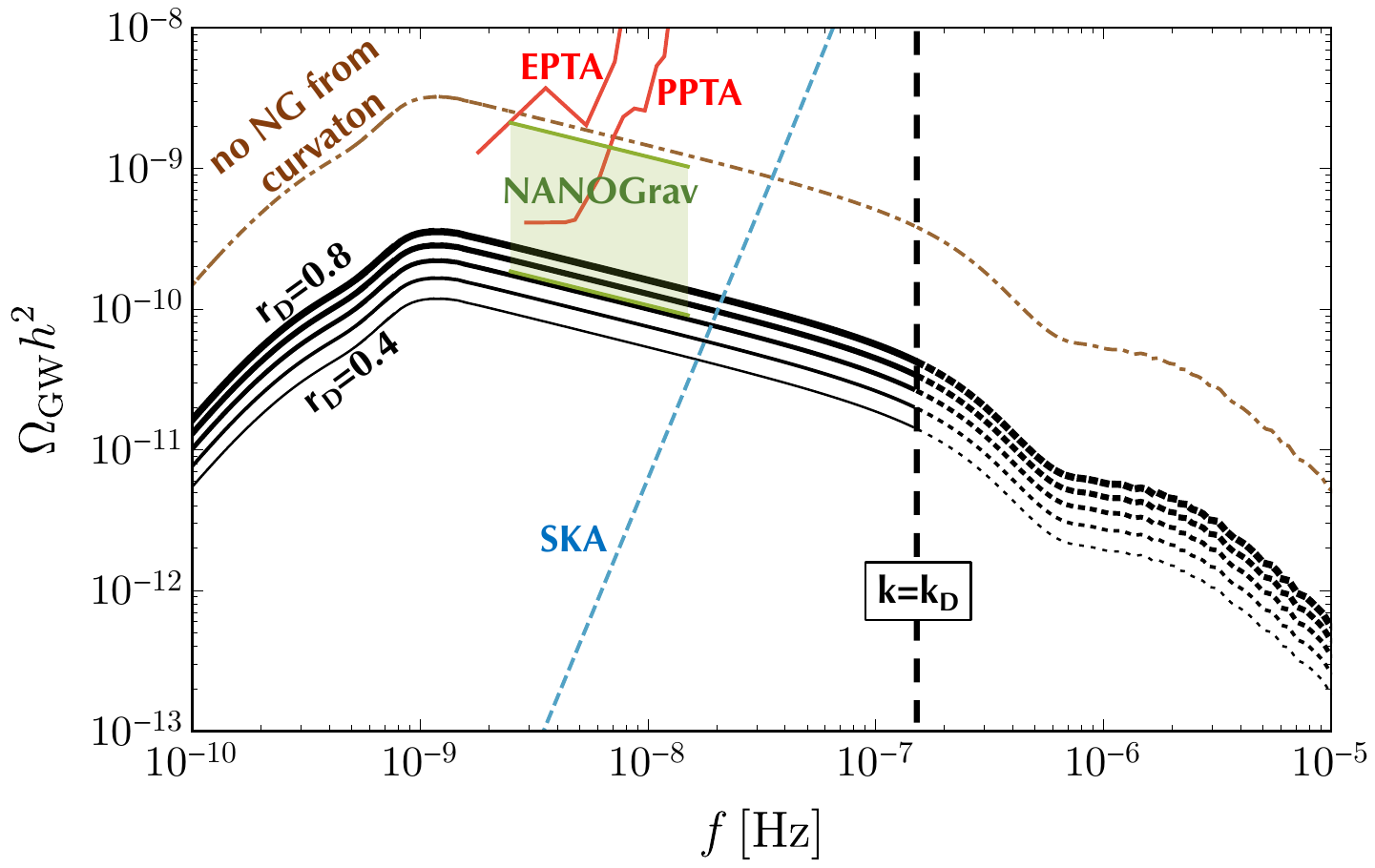}
        \caption{
        The spectra of the induced GWs with $r_\DD=0.8$ (black, top, thick) $\sim$ $0.4$ (bottom, thin).
        The difference of $r_\DD$ between two adjacent lines is $0.1$.
        $p(0.8) = 4.4$ is taken for the coefficient of the skewness given in Eq.~(\ref{eq:skewness_exp}) because of $n_2 = 0.8$.
        We also show the spectrum without the non-Gaussianity from curvaton ($\mu_3/\sigma = -9/4$, brown dot-dashed).
        The green-shaded region shows the 2$\sigma$ region of the NANOGrav results with the tilt of $\Omega_\text{GW} \propto f^{-0.4}$.
        The constraints are also shown from the previous results of EPTA~\cite{Lentati:2015qwp} and PPTA~\cite{Shannon:2015ect}, and the future sensitivity of SKA~\cite{Janssen:2014dka}.
        The black dotted lines show the spectra on $k>k_\DD$, which might be suppressed more because the curvaton decay is not instantaneous (see below Eq.~(\ref{eq:f_def})). 
        }
        \label{fig:gw_spectrum}
\end{figure}

\noindent
Figure~\ref{fig:pbh_mass_spectrum} shows the PBH mass spectrum with this parameter set. 
We can see $f_\text{PBH} \sim 10^{-3}$ at $M = 30 M_\odot$, which is consistent with the merger rate estimated by the LIGO-Virgo collaboration~\cite{Sasaki:2016jop,Sasaki:2018dmp}\footnote{
Refs.~\cite{Raidal:2018bbj,DeLuca:2020qqa} discuss how this estimation of $f_\text{PBH} \sim 10^{-3}$ gets modified by the actual shape of the PBH mass spectrum. 
Since the PBH mass spectrum is exponentially sensitive to the curvature perturbations and hence the induced GWs, we may easily accommodate this change by a small modification of the model parameters while having GWs consistent with the NANOGrav result.
Also, there are several uncertainties in the computation of the PBH mass spectrum for given curvature perturbations, for instance \cite{Young:2019osy,Ando:2018qdb}.
For these reasons, we just take $f_\text{PBH} \sim 10^{-3}$ as a representative value.
}.
The bump around $M \sim 1 M_\odot$ is due to the suppression of the PBH threshold during the QCD phase transition~\cite{Byrnes:2018clq}.
Although the mass spectrum seems inconsistent with the constraint from the CMB anisotropy in Ref.~\cite{Serpico:2020ehh}, we should keep in mind that it still has uncertainties in accretion models, as the reference itself mentions.

Figure~\ref{fig:gw_spectrum} shows the spectra of GWs induced by the scalar perturbations that realize the mass spectrum in Fig.~\ref{fig:pbh_mass_spectrum}. 
Note again that the non-Gaussianity, dependent on $r_\DD$, changes the relation between the abundance of PBHs and the power spectrum of curvature perturbations.
From this figure, we can see that if we take $r_\text{D} \gtrsim 0.6$, the induced GWs are consistent with the 2$\sigma$ region of the NANOGrav stochastic process.
On the other hand, for a small $r_\DD \lesssim 0.6$, corresponding to $f_\text{NL} \gtrsim 2.1$, the non-Gaussianity suppresses the induced GWs too much to explain the  NANOGrav results.
For comparison, we also provide the induced GWs neglecting the primordial non-Gaussianity from the curvaton.
We can see that, if there is no primordial non-Gaussianity, the induced GWs are too large to be compatible with the NANOGrav results.
This is qualitatively consistent with Ref.~\cite{Vaskonen:2020lbd}.

Regarding the parameters except for $k_\text{D}$, we cannot take them much differently from the values in Eq.~(\ref{eq:parameter_set}) to be consistent with the LIGO-Virgo observation and the NANOGrav result. Only the $k_\DD$ is independent of the observation results and we can change it freely as long as $k_\DD$ is larger than the NANOGrav frequency range. Note again that the GW spectrum on the frequency corresponding to $k>k_\text{D}$ has the uncertainty (see below Eq.~(\ref{eq:f_def})).

\emph{Conclusion.}---
In this letter, we discuss the implications of the recent NANOGrav results in terms of PBHs that explain the BHs detected by the LIGO-Virgo collaboration.
Such PBHs for the LIGO-Virgo events predict large GWs, which are induced by scalar perturbation, around the NANOGrav frequency.
We pursue the possibility to explain the NANOGrav results by these induced GWs.
We study the axion-like curvaton model as a concrete example, which inevitably produces the primordial non-Gaussianity.
We have demonstrated that the curvaton model can account for the LIGO-Virgo PBH scenario and the NANOGrav results at the same time, which is enabled
by the non-Gaussianity suppressing the induced GWs with the PBH abundance fixed.
Our result implies that the NANOGrav results could be a signal from the LIGO-Virgo PBHs in the case where the primordial non-Gaussianity exists.

{ 
\noindent
\emph{Acknowledgments.}---
K.I. thanks Kenta Ando for useful comments on the manuscript.
T.T.Y. thanks Satoshi Shirai for useful discussion on effects from the QCD phase transition.
This work was supported in part by 
the Kavli Institute for Cosmological Physics at the University of Chicago through an endowment from the Kavli Foundation and its founder Fred Kavli (K.I.), 
World Premier International Research Center Initiative (WPI Initiative), MEXT, Japan,
JSPS KAKENHI Grants No.~17H01131 (M.K.), No.~17K05434 (M.K.), 
No.~17H02878 (T.T.Y.), and No.~19H05810 (T.T.Y.), 
MEXT KAKENHI Grant No.~15H05889 (M.K.),
the China Grant for Talent Scientific Start-Up Project (T.T.Y.), and 
the Deutsche Forschungsgemeinschaft under Germany's Excellence Strategy - EXC 2121 Quantum Universe - 390833306 (K.M.).
}

\small
\bibliographystyle{apsrev4-1}
\bibliography{nanograv_pbh}

\end{document}